\documentclass[12pt,preprint]{aastex}

\shorttitle{HD 101088, AN ACCRETING 14 AU BINARY IN LCC}
\shortauthors{BITNER ET AL.}

\begin{document}

\title{HD 101088, AN ACCRETING 14 AU BINARY IN LOWER CENTAURUS CRUX WITH VERY LITTLE CIRCUMSTELLAR DUST\footnote{This paper 
includes data gathered with the 6.5 meter Magellan Telescopes located at Las Campanas Observatory, Chile.}}
\author{Martin A. Bitner\altaffilmark{a}, Christine H. Chen\altaffilmark{a}, James Muzerolle\altaffilmark{a}, Alycia J. Weinberger\altaffilmark{b}, Mark Pecaut\altaffilmark{c}, Eric E. Mamajek\altaffilmark{c},Melissa K. Mclure\altaffilmark{d}}

\altaffiltext{a}{Space Telescope Science Institute, Baltimore, MD; mbitner@stsci.edu}
\altaffiltext{b}{Department of Terrestrial Magnetism, Carnegie Institution of Washington, Washington, DC}
\altaffiltext{c}{University of Rochester, Department of Physics and Astronomy, Rochester, NY}
\altaffiltext{d}{University of Michigan, Department of Astronomy, Ann Arbor, MI}

\begin{abstract}

We present high resolution (R=55,000) optical spectra obtained with MIKE on the 6.5 m 
Magellan Clay Telescope as well as \textit{Spitzer} MIPS photometry and IRS low resolution (R$\sim$ 60) spectroscopy 
of the close (14 AU separation) binary, HD 101088, a member of the $\sim$12 Myr old southern region of the 
Lower Centaurus Crux (LCC) subgroup of the Scorpius-Centaurus OB association.
We find that the primary and/or secondary is accreting from a tenuous circumprimary and/or circumsecondary 
disk despite the apparent lack of a massive circumbinary disk.
We estimate a lower limit to the accretion rate of \.M $> 1\times10^{-9} ~M_{\odot} ~yr^{-1}$, which our 
multiple observation epochs show varies over a timescale of months.
The upper limit on the 70 $\mu$m flux allows us to place an upper limit on the mass of dust grains 
smaller than several microns present in a circumbinary disk of 0.16 M$_{moon}$. 
We conclude that the classification of disks into either protoplanetary or debris disks based on fractional 
infrared luminosity alone may be misleading.
\end {abstract}


\section{INTRODUCTION}
Most young stars are initially surrounded by optically thick accretion disks \citep{beckwith90}.
Though the frequency of binarity is a function of mass and formation environment, one-third of 
all main sequence stars in the Galactic disk are in binaries \citep{lada06}. 
In nearby star forming regions, more than half \citep{ghez93,simon95} of young stars have been observed 
to be members of binary systems. 
Since the formation of a binary is a common outcome of the star formation process, 
studying the structure and evolution of disks in binary systems is an important part of 
developing a complete understanding of the planet formation process.
In the case of young binary systems, disks may surround the primary, the secondary, and/or 
both components \citep{artymowicz94}.
The separation between stars in a binary system is an important parameter that affects 
the geometry and structure of any disk material surrounding the stars.
Optically thick disks around each member of young binaries have been observed 
for stars separated by as little as 14 AU \citep{hartigan03}. 
The outer radii of disks in close binary systems are truncated through gravitational interactions between 
the binary components, limiting the amount of material available for planet formation and accretion onto the stars.

Optically thick accretion disks around single stars appear to dissipate within a few million years 
\citep{haisch01}.
Since the disks around each star in close binary systems are truncated to smaller outer radii 
than disks around single stars, they may disperse faster.
\citet{bouwman06} conducted a survey with the \textit{Spitzer Space Telescope} of the $\sim$8 Myr old $\eta$ 
Chamaeleontis cluster and found that circumstellar disks were detected around 80\% of single 
stars yet absent around 80\% of the close binary stars.
This is suggestive of a shorter timescale for disk removal in close binaries although the 
sample size was small and the binaries were not spatially resolved.
Additional observations of disks in close binary systems are needed to confirm these results.
  
Here we present Magellan Inamori Kyocera Echelle (MIKE) R=55,000 optical spectroscopy 
along with \textit{Spitzer} Infrared Spectrometer (IRS; Houck et al. 2004) and 
Multiband Imaging Photometer for \textit{Spitzer} (MIPS; Rieke et al. 2004) observations 
of the close binary star, HD 101088, a member of the $\sim$12 Myr old southern region of the Lower Centaurus-Crux (LCC) 
subgroup of the Scorpius-Centaurus OB association \citep{dezeeuw99}.
HD 101088 consists of a F5 primary star \citep{houk75} and a secondary of unknown spectral type according 
to \textit{Hipparcos} astrometry. 
The two components are separated by 0.15\arcsec or 14 AU at a distance of 94 pc \citep{vanleeuwen07}.
Our high resolution optical spectra reveal broad, spatially unresolved H$\alpha$ emission from this source which 
is indicative of ongoing stellar accretion.
The \textit{Spitzer} IRS spectrum shows there is very little, if any, hot circumstellar dust.
The lack of strong emission in the mid- and far-infrared indicates the absence of a cold outer disk.

\section{OBSERVATIONS AND DATA REDUCTION}

We observed HD 101088 with the MIKE spectrograph \citep{bernstein03} on the 6.5 m Magellan Clay Telescope at 
Las Campanas Observatory on 2007 March 11, 2009 April 15, and 2009 June 8 (UT).
The 0.35\arcsec $\times$ 5\arcsec ~slit was used in all cases, giving a resolution 
of 55,000 at the wavelength of H$\alpha$.
The spectra were flat-fielded, extracted and wavelength calibrated using the 
MIKE pipeline written by D. Kelson with techniques described in 
\citet{kelson00}, \citet{kelson03}, and \citet{kelson06}. 
The exposure times were 24, 60, and 147 seconds for the 2007 March, 2009 April, and 2009 June observations respectively.
The resulting spectrum has a signal-to-noise ratio larger than 100 per pixel for wavelengths greater than 4500 \AA.

Photometry at 24 and 70 $\mu$m was obtained using MIPS on \textit{Spitzer}.
The observations were made on 2005 April 9 with integration times of one cycle of 3 s at 24 $\mu$m 
and one cycle of 10 s at 70 $\mu$m.
The Data Analysis Tool (DAT), version 2.80, created by the MIPS instrument team \citep{gordon05} 
was used to reduce the data.
The MIPS calibration uncertainty was taken to be 4\% at 24 $\mu$m and 7\% at 70 $\mu$m based on the MIPS handbook.
Additional MIPS data processing details can be found in C. Chen et al. (2010, in preparation).
A \textit{Spitzer} IRS spectrum was obtained on 2007 June 18 with both the 
Short-Low (5.2-14.0 $\mu$m) and the Long-Low (14.0-38.0 $\mu$m) modules.
The observations were carried out in IRS staring mode with no peak-up.
The Short-Low observations consisted of 2 cycles of 6 seconds each while the Long-Low observations 
were 2 cycles of 14 seconds.
The data were reduced and analyzed using the SMART program created by the IRS team \citep{higdon04} 
following the procedures described in \citet{furlan06}, with sky subtraction from the opposite nod 
position.
We estimate the spectrophotometric uncertainty of the IRS spectrum to be approximately 5\%.

\section{RESULTS}
We estimated the age of the system by placing the star on the HR diagram and comparing to 
pre-main-sequence evolutionary tracks and by considering the age of the ensemble of young stars 
with which HD 101088 is associated.
Isochronal age estimates for HD 101088 range from 2 Myr using the tracks of \citet{palla01} to 5 Myr based on the \citet{siess2000} tracks.
Since observational uncertainties can lead to errors in age estimates for individual stars, we also considered 
the average age of the region of Lower Centaurus Crux where HD 101088 is located.
\citet{preibisch08} found an average age for the members of the southern portion of Lower Centaurus Crux 
of 12 Myr so we conservatively estimate the actual age of HD 101088 as being between 2 and 12 Myr.

The projected rotational velocity ($v$ sin $i$) was determined by fitting rotationally broadened 
synthetic spectra to our data.
The fitting region was restricted to 4000-7000 \AA, excluding regions with strong 
telluric contamination, H$\alpha$, and the Na D lines.
We used Richard Gray's spectral synthesis program, SPECTRUM\footnote{http://www.phys.appstate.edu/spectrum/spectrum.html}, along with Kurucz model atmospheres of solar metallicity 
to compute the synthetic spectra and then broadened them using the rotational profile 
given in \citet{gray92}.
We fitted these synthetic spectra to our data and performed a $\chi^{2}$ minimization to find the best value 
for $v$ sin $i$.
The average of the measurements derived from all the separate observations is listed in Table~\ref{tab:props}.
The RVCORRECT and FXCOR packages in IRAF were used to measure the radial velocity.
Measurements from the three different observations were consistent with one another and the average 
is 17.8 $\pm$ 3.9 km s$^{-1}$.
By combining the position, proper motion, and parallax of HD 101088 from the revised \textit{Hipparcos} catalog \citep{vanleeuwen07} 
with our measured radial velocity, we derive a Galactic space velocity (U,V,W) = -5.6 $\pm$ 1.9, -21.9 $\pm$ 2.8, -6.5 $\pm$ 2.1 
km s$^{-1}$, consistent with the mean space motions for members of Lower Centaurus Crux (E. Mamajek, private communication).

The spectra from all observed epochs show broad H$\alpha$ emission as seen in Figure~\ref{fig:halpha}.
In order to characterize the H$\alpha$ emission which falls on top of photospheric 
absorption, we subtracted a broadened HD 106444 spectrum, a star with 
the same spectral type as the HD 101088 primary and somewhat smaller $v$ sin $i$ (96 km s$^{-1}$ vs. 160 km s$^{-1}$) 
which we also observed using MIKE.
The H$\alpha$ measurements we report are based on the subtracted spectra.
The H$\alpha$ emission has an equivalent width of 3.6, 4.3, and 6.2 \AA ~in 2007 March, 2009 April, 
and 2009 June respectively.
We estimate the systematic uncertainty for the equivalent width measurements to be 0.4 \AA which is the standard 
deviation of the measured equivalent widths of the H$\alpha$ absorption in 14 F5 stars we have observed in Sco Cen.  
The H$\alpha$ full width at 10\% of the peak is commonly used to differentiate 
between accretion and chromospheric emission.
H$\alpha$ 10\% widths $>$ 270 km s$^{-1}$ are due to accretion independent of spectral type \citep{white03}.   
For HD 101088, the H$\alpha$ 10\% full width is broad but variable over our three different observations; 388 $\pm$ 2, 
380 $\pm$ 2, and 429 $\pm$ 4 km s$^{-1}$ in 2007 March, 2009 April, and 2009 June respectively, suggesting the presence 
of ongoing accretion.

We compared the observed H$\alpha$ profiles with radiative transfer models
of magnetospheric accretion \citep{muzerolle01}.  These models
are consistent with the line emission observed in most classical T Tauri stars,
as well as some Herbig Ae stars \citep{muzerolle04}, and so provide
potential constraints on the accretion activity in the HD 101088 system.
We calculated models using the mass, radius, and effective temperature of
the primary star as fixed inputs.  The outer radius of the magnetosphere
was fixed at the corotation radius, which is about 1.5 R$_{*}$ given the $v$ sin $i$
value we measure for the primary.  The gas temperature, density, and
inclination were then varied to find a good fit to the observed profile.
Figure~\ref{fig:halpha} shows one example fit to the 2007 observation. 
We are also able to fit the profile assuming
that the secondary rather than the primary is accreting, using similar
parameters and accounting for the additional continuum from the unresolved
primary.  Either model has difficulty matching the increased line emission
in the 2009 spectrum; the accretion geometry may have changed, or
an additional source of emission such as a wind may have manifested itself. 
The difficulty in fitting the June 2009 profile is that the models 
do not produce sufficiently strong emission for any reasonable set of parameters.
The only way to increase the emission would be to increase the size of the magnetosphere, 
but that is not possible since it cannot be larger than the corotation radius.
One way around this would be to invoke a different geometry from the dipole approximation, 
such as a pinched or inflated field configuration.
Note that because of degeneracy in constraining the gas temperature and
density, we can only put a lower limit to the actual accretion rate,
\.M $> 1\times10^{-9} ~M_{\odot} ~yr^{-1}$.  
Spatially resolved observations of other diagnostics,
such as infrared emission lines or UV continuum excess, are needed to obtain
better constraints on the accretion rate and its origin in the system. 

The \textit{Spitzer} MIPS photometry gives fluxes of 70.0 $\pm$ 2.2 mJy at 24 $\mu$m and 
a 3-$\sigma$ upper limit of 16.5 mJy at 70 $\mu$m.
The uncertainty for the 24 $\mu$m measurement was computed by adding the 0.75 mJy statistical error 
to the 3\% calibration uncertainty in quadrature.
The primary star in HD 101088 is known to have a spectral type of F5 \citep{houk75}.
In order to compute the spectral energy distribution for the stars in HD 101088 to look for an infrared excess, 
we need to determine the spectral type of the secondary star.
\textit{Hipparcos} gives H$_{p}$=6.8$\pm$0.07 for the primary and H$_{p}$=9.6$\pm$0.89 for the secondary.
Assuming an age of 5 Myr and knowing that the spectral type of the primary is F5, we used the \citet{siess2000} 
evolutionary tracks to translate the magnitude difference between the two stars, including uncertainties, 
to a range for the secondary's spectral type of K0-K5.
We computed the stellar photospheric flux by normalizing Kurucz model atmospheres for both components of the binary 
to \textit{Hipparcos} photometric values.
We assume HD 101088 has solar metallicity and our calculated value for the line-of-sight extinction of A$_{v}$=0.13.
Figure~\ref{fig:photfit} shows the combined stellar photospheric flux plotted over fluxes at B and V from 
\textit{Hipparcos} and J, H, and K from 2MASS.
The upper panel shows the SED assuming a K5 secondary while the lower panel shows the SED in the case of a K0 secondary 
overplotted with the IRS spectrum, MIPS 24 $\mu$m point, and MIPS 70 $\mu$m upper limit.
The IRS spectrum appears to have nearly the same slope as the stellar photosphere and shows no silicate emission features.
In the case of a K5 secondary, the MIPS 24 $\mu$m flux is 6.2$\sigma$ larger than the 56.4 mJy flux from the combined stellar 
photospheres.
We modeled the infrared excess emission assuming that the dust grains are large, consistent with the 
featureless \textit{Spitzer} IRS spectrum. 
However, if dust grains smaller than a few microns are present, the lack of spectral features 
in the IRS spectrum requires that the dust grains are composed of material other than silicates, 
perhaps carbon or iron.
Because we lack measurements at wavelengths between the 2MASS photometric points and the IRS spectrum, 
acceptable fits can be achieved for a wide range of dust temperatures, 500-2000 K. 
If the secondary is a K0 star, then the MIPS 24 $\mu$m flux is only 1.7$\sigma$ larger than the 66.2 mJy combined stellar 
photospheric flux and consistent with there being no infrared excess.
However, with a K0 secondary star, the SED overpredicts the flux at J, H, and K by 8.4$\sigma$, 12.1$\sigma$, and 4.1$\sigma$ 
respectively.
A K1 spectral type is the hottest allowable secondary star which agrees with the J, H, and K photometry.
Given the uncertainty about the nature of the secondary star in the system, we cannot determine whether 
there is an infrared excess due to any warm circumstellar dust.
After subtracting a F5 template from our HD 101088 spectra, we searched for any leftover absorption lines 
due to the secondary star and found none.
This does not necessarily rule out a companion since they could be heavily veiled by light from the primary 
or any accretion excess from the secondary if it is accreting.
Or, perhaps they are broadened due to rapid rotation.
Spatially resolved observations are needed to characterize the spectral type of the secondary star.

The \textit{Spitzer} MIPS 70 $\mu$m upper limit can be used to place a constraint on the 
presence of any cold dust in the system.
An upper limit on the mass of circumstellar dust in the system contributing to 
the 70 $\mu$m flux can be made assuming the dust is optically thin and at a single temperature.
In that case, the dust mass is given by $M_{dust}=\frac{F_{\nu}D^{2}_{*}}{B_{\nu}(T)\kappa_{\nu}}$ \citep{jura95}, 
where $D_{*}$ is the distance to the star and $\kappa_{\nu}$ is the dust absorption opacity.
If we assume a circumbinary disk is present and take the semi-major axis of the binary orbit to be 
half of the \textit{Hipparcos} separation between binary components, 
the inner radius of the circumbinary disk would be at 15 AU or 2.08 
times the semi-major axis of the binary orbit, the expected location of the inner edge of a circumbinary 
disk for a binary with circular orbits \citep{artymowicz94}.
We estimate a temperature of $T=150$K at this location by simple radiative balance with the stars, 
$T\sim (L_{*}/16 \pi r^{2}\sigma)^{1/4}$, where $\sigma$ is the Stefan-Boltzmann constant.
With $F_{\nu}$, the upper limit on the 70 $\mu$m flux equal to 16.5 mJy, the distance 
to HD 101088 taken to be 94 pc, and $\kappa_{\nu} = 3$ cm$^{2}$ g$^{-1}$ \citep{pollack94}, 
$M_{dust} < 1.2\times 10^{25}$g or 0.16 M$_{moon}$.
If dust grains much larger than several microns are present in the outer disk, the amount of mass 
there could be substantially larger.

\section{DISCUSSION}
The presence of accretion in the absence of a large infrared excess is surprising.
Theoretical models have shown that circumstellar disks in close binary systems are 
expected to be tidally truncated to an outer radius significantly smaller than disks around single stars 
\citep{artymowicz94}.
They give the expected truncation radii for circumprimary and circumsecondary 
disks based on the semi-major axis of the binary orbit.
Assuming circular orbits, the outer disk radii are 0.46a and 0.2a for circumprimary and 
circumsecondary disks respectively.
For HD 101088, this corresponds to outer disk radii of 6.5 AU and 3 AU.

Since our observations do not spatially resolve the two binary components, we cannot decipher 
which of the two stars has a disk or if both do.
However, in either case, the timescale to viscously accrete the material from such tidally 
truncated disks is very short.
We follow the viscous accretion calculations in \citet{quillen04} to compute the time required 
to deplete the disk material.
The accretion timescale is given by $\tau_{\nu}=\alpha^{-1}(\frac{r}{h})^{2}\tau_{orb}/2\pi$, 
where $\alpha$ is the viscosity parameter, $r$ is the disk outer radius, $h$ is the disk 
scale height, and $\tau_{orb}$ is the orbital period of material at the outer disk radius.
We take $\alpha$ to be 0.01, a value typical for accretion disks, and $h/r$ is estimated from 
the parametrization given by \citet{chiang97}, $h/r=0.17a_{AU}^{2/7}$.
For the truncated circumprimary disk, the orbital period at the disk truncation radius is 14 years and 
the time to clear the disk through viscous accretion is 2650 years.
The slightly smaller circumsecondary disk would accrete all its material in only 1800 years.

Both timescales are much shorter than the age of the system implying that the circumstellar 
material must have been replenished, perhaps from a circumbinary disk.
There are examples of strongly accreting stars which appear to be devoid of material to accrete 
based on the lack of near-infrared excess emission such as V4046 Sgr \citep{jensen97}, however, in 
contrast to HD 101088, evidence is seen of an outer disk reservoir in the form of large excess emission 
at wavelengths longward of 10 $\mu$m. 
If a reservoir of material exists in a circumbinary disk, material may be transferred to a circumstellar disk 
around one of the stars \citep{artymowicz96}.
In this model, circumbinary disks can transfer material via pulsed accretion to the circumstellar disks 
of the individual stars to support accretion rates comparable to that found for single stars with 
full disks.
Some observational evidence for pulsed accretion which varies as a function of orbital phase 
has been found.
\citet{jensen07} reported BVRI photometry of the pre-main-sequence spectroscopic binary UZ Tau E 
which varied on a timescale consistent with the period of the binary suggesting that periodic accretion 
can occur from a circumbinary disk.
Variability in the H$\alpha$ emission was not as clearly periodic, perhaps due to a lack of data.
The variation we observe in the HD 101088 H$\alpha$ emission over a timescale of a few months is unlikely 
to be due to pulsed accretion from a circumbinary disk since the orbital period of the system, $\sim$35 
years, is significantly longer. 

It is not clear from our observations that HD 101088 possesses a circumbinary disk from which to 
replenish material around the individual stars.  
Our \textit{Spitzer} observations suggest there is very little material that could be present in a circumbinary disk.
The dust mass estimate based on the upper limit at 70 $\mu$m of $<0.16$ M$_{moon}$ is 
eight orders of magnitude smaller than the 0.13 M$_{\odot}$ circumbinary disk around GG Tau \citep{dutrey94}.
A larger survey of circumbinary disk masses has found typical upper limits of 0.005 M$_{\odot}$ for binaries 
with separations less than 100 AU \citep{jensen96}, still significantly larger than our derived upper limit for HD 101088. 
Our observations suggest there is no circumbinary disk present and we have caught 
HD 101088 just as it is clearing its remaining circumstellar material.
Spatially resolved observations to determine which star is accreting and submillimeter data to probe any outer disk 
material present are needed to further elucidate the properties of this interesting system.

\section{CONCLUSIONS}
We have obtained high resolution optical spectroscopy with MIKE on the Magellan Clay telescope as well as 
\textit{Spitzer} IRS spectroscopy and MIPS 24,70 $\mu$m photometry of the close binary, 
HD 101088. 

1. Broad H$\alpha$ emission is present in our spectra and reveals ongoing 
stellar accretion and the presence of circumstellar gas.  We derive a lower limit on the accretion rate of 
\.M $> 1\times10^{-9} ~M_{\odot} ~yr^{-1}$.  

2. The truncated disks in such a close binary have viscous accretion lifetimes much shorter 
than the age of the system, suggesting that some source of replenishment must be present 
to maintain the ongoing accretion.

3. The upper limit at 70 $\mu$m leads to a constraint on the dust mass in a circumbinary disk if it 
is present of $<0.16$ M$_{moon}$.

4. HD 101088 would be classified as having a debris disk based on its small 
fractional infrared luminosity, 7.0$\times$10$^{-4}$.  The presence of ongoing 
accretion shows that the classification of disks into either protoplanetary 
or debris disks based on fractional infrared luminosity alone may be misleading.

5. We find that the primary and/or secondary is accreting from a tenuous circumprimary 
and/or circumsecondary disk despite the apparent lack of a massive circumbinary disk.  
This unique situation merits further study. 

\acknowledgements
We thank E. Shkolnik for obtaining a spectrum of HD 101088 in June 2009.
This research has made use of the SIMBAD database, operated at CDS, Strasbourg, France.

\figcaption[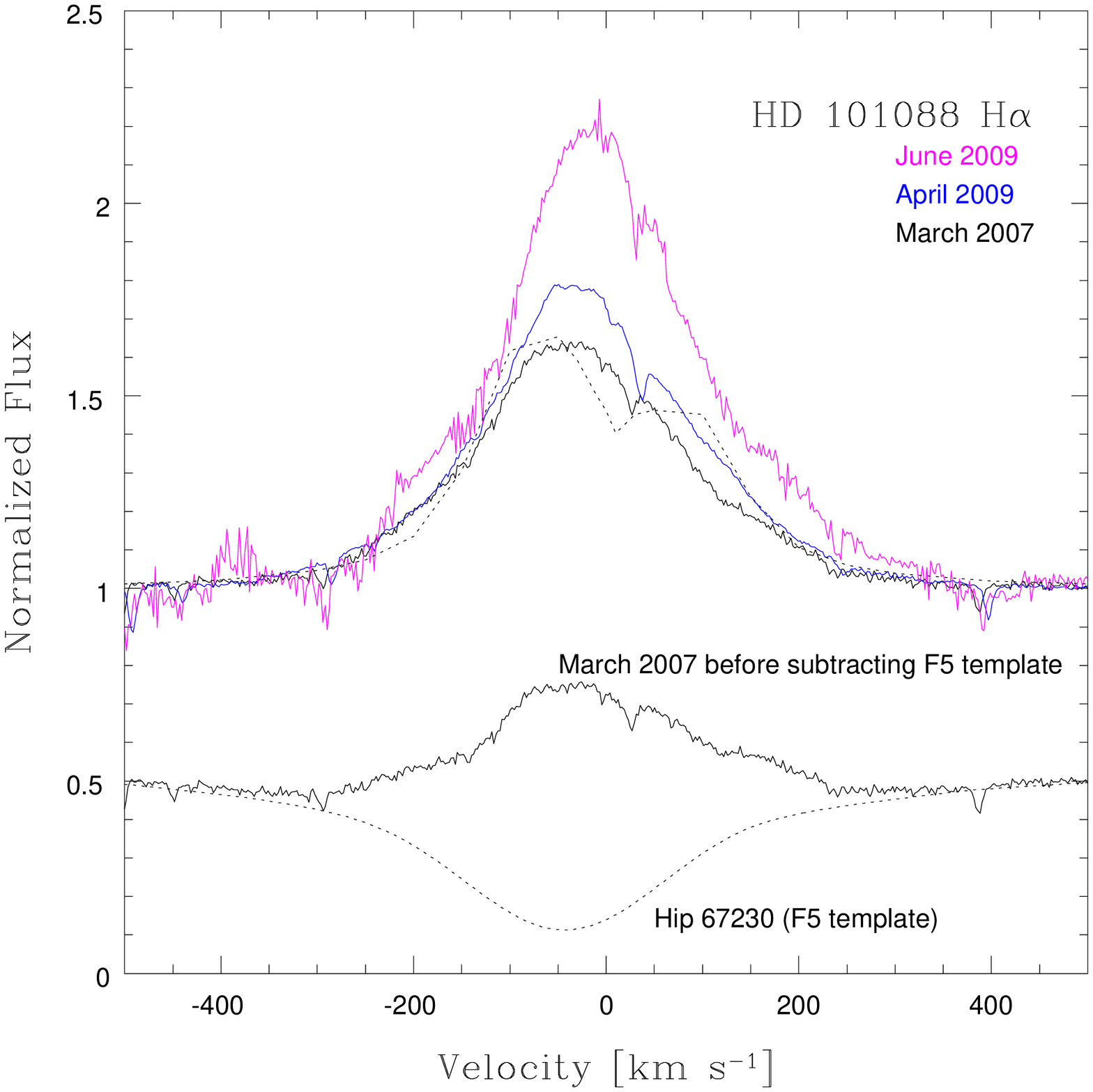]
{Three H$\alpha$ line profiles obtained during different epochs of observation overplotted 
with an example accretion model line profile (dashed line) for the March 2007 observation.
The parameters for the model line profile are \.M $=1\times 10^{-8} ~M_{\odot} ~yr^{-1}$, T$_{max}$=10,000 K,
\textit{i}=10$^{\circ}$, and R$_{mag}$=1.3-1.5 R$_{*}$.
Due to degeneracy in constraining the gas temperature and
density, we can only put a lower limit to the actual accretion rate,~\.M $> 1\times10^{-9} ~M_{\odot} ~yr^{-1}$.  
The raw March 2007 spectrum around H$\alpha$ before subtracting the template spectrum to remove the underlying photospheric 
absorption is shown offset from the three line profiles.   
\label{fig:halpha}
}

\figcaption[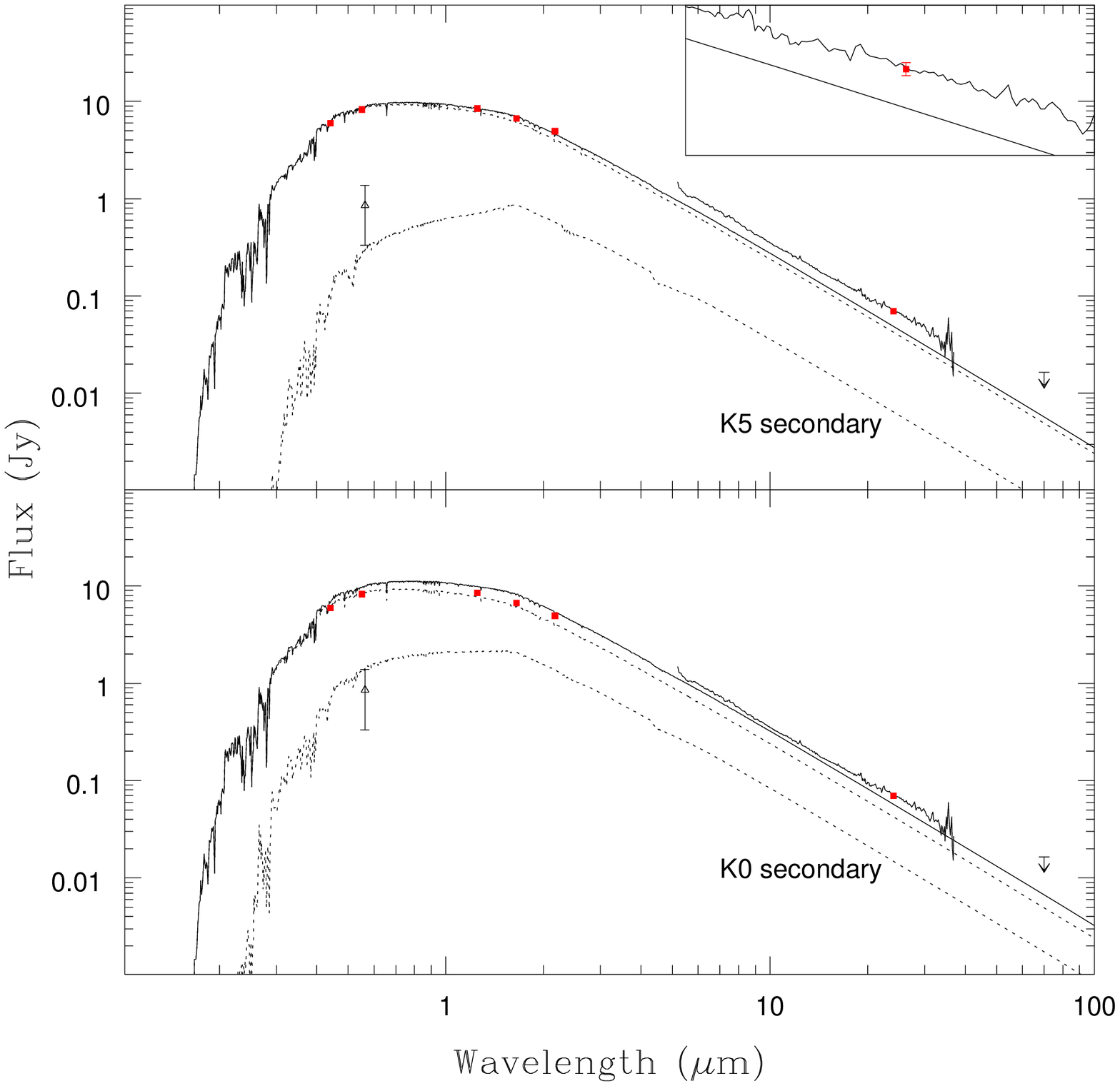]
{Possible SEDs for the range of secondary spectral types allowed by \textit{Hipparcos} observations 
and uncertainties. 
The top panel assumes a K5 secondary and shows the total contribution from both components of the HD 101088 binary plotted 
over B and V fluxes from the Tycho-2 Catalogue and 2MASS J,H, and K fluxes.
The error bars on the B,V,J,H, and K values are smaller than the plotted symbols.
The Hipparcos H$_{p}$ magnitude of the secondary star is plotted along with its error bar.
The dashed lines show the individual contributions from the two stars, a F5 primary and K5 secondary.
The \textit{Spitzer} IRS spectrum is shown along with the MIPS 24 $\mu$m photometric point and MIPS 70 $\mu$m 
3$\sigma$ upper limit.
The inset shows a region around the MIPS 24 $\mu$m point with the 1$\sigma$ error bar overplotted.
The combined photospheric flux in this case is 6.2$\sigma$ below the MIPS 24 $\mu$m point. 
The bottom panel shows the same with a K0 secondary.
In this case, the combined photospheric flux is only 1.7$\sigma$ below the MIPS 24 $\mu$m point but 
above the 2MASS J,H,K points by 8.4$\sigma$, 12.1$\sigma$, and 4.1$\sigma$ respectively.
\label{fig:photfit}
}

\clearpage
\begin{figure}
\figurenum{\ref{fig:halpha}}
\plotone{f1.eps}
\end{figure}

\clearpage
\begin{figure}
\figurenum{\ref{fig:photfit}}
\plotone{f2.eps}
\end{figure}

\clearpage
\begin{deluxetable}{lcc}
\tablewidth{4.0in}
\tablecaption{Properties of HD 101088\label{tab:props}}
\tablehead{
\colhead{Property} &
\colhead{Value} &
\colhead{Reference}\\
}

\startdata
Primary Spectral Type & F5IV & 1\\
Secondary Spectral Type & K0-K5\tablenotemark{a} & 2 \\
Distance (pc)& 94$^{+5}_{-4}$ & 3 \\
Age (Myr) & 2-12\tablenotemark{b} & \nodata \\
RV (km s$^{-1}$) & 17.8 $\pm$ 3.9 & 2\\ 
$v$ sin $i$ (km s$^{-1}$) & 160 $\pm$ 4 & 2\\
H$\alpha$ EW (\AA) & 3.6/4.3/6.2\tablenotemark{c}\tablenotemark{d} & 2\\
H$\alpha$ 10\% FW (km s$^{-1}$) & 388/380/429\tablenotemark{c}\tablenotemark{e} & 2\\
L$_{IR}$/L$_{*}$ & 7.0$\times$10$^{-4}$ & 2\\
\enddata

\tablenotetext{a}{Spectral type range based on \textit{Hipparcos} magnitude and uncertainties. }
\tablenotetext{b}{2-5 Myr from HR diagram fits, 12 Myr average age of southern LCC}
\tablenotetext{c}{11mar07/15apr09/08jun09 }
\tablenotetext{d}{H$\alpha$ EW Systematic Uncertainty: 0.4 \AA}
\tablenotetext{e}{H$\alpha$ 10\% FW Uncertainties: 2/2/4 km s$^{-1}$}
\tablerefs{(1) \citet{houk75}. ~(2) this work. ~(3) \citet{vanleeuwen07}.}

\end{deluxetable}


\begin{thebibliography}{36}
\expandafter\ifx\csname natexlab\endcsname\relax\def\natexlab#1{#1}\fi

\bibitem[{{Artymowicz} \& {Lubow}(1994)}]{artymowicz94}
{Artymowicz}, P., \& {Lubow}, S.~H. 1994, \apj, 421, 651

\bibitem[{{Artymowicz} \& {Lubow}(1996)}]{artymowicz96}
---. 1996, \apjl, 467, L77+

\bibitem[{{Beckwith} {et~al.}(1990){Beckwith}, {Sargent}, {Chini}, \&
  {Guesten}}]{beckwith90}
{Beckwith}, S.~V.~W., {Sargent}, A.~I., {Chini}, R.~S., \& {Guesten}, R. 1990,
  \aj, 99, 924

\bibitem[{{Bernstein} {et~al.}(2003){Bernstein}, {Shectman}, {Gunnels},
  {Mochnacki}, \& {Athey}}]{bernstein03}
{Bernstein}, R., {Shectman}, S.~A., {Gunnels}, S.~M., {Mochnacki}, S., \&
  {Athey}, A.~E. 2003, in Society of Photo-Optical Instrumentation Engineers
  (SPIE) Conference Series, Vol. 4841, Society of Photo-Optical Instrumentation
  Engineers (SPIE) Conference Series, ed. M.~{Iye} \& A.~F.~M. {Moorwood},
  1694--1704

\bibitem[{{Bouwman} {et~al.}(2006){Bouwman}, {Lawson}, {Dominik}, {Feigelson},
  {Henning}, {Tielens}, \& {Waters}}]{bouwman06}
{Bouwman}, J., {Lawson}, W.~A., {Dominik}, C., {Feigelson}, E.~D., {Henning},
  T., {Tielens}, A.~G.~G.~M., \& {Waters}, L.~B.~F.~M. 2006, \apjl, 653, L57

\bibitem[{{Chiang} \& {Goldreich}(1997)}]{chiang97}
{Chiang}, E.~I., \& {Goldreich}, P. 1997, \apj, 490, 368

\bibitem[{{de Zeeuw} {et~al.}(1999){de Zeeuw}, {Hoogerwerf}, {de Bruijne},
  {Brown}, \& {Blaauw}}]{dezeeuw99}
{de Zeeuw}, P.~T., {Hoogerwerf}, R., {de Bruijne}, J.~H.~J., {Brown}, A.~G.~A.,
  \& {Blaauw}, A. 1999, \aj, 117, 354

\bibitem[{{Dutrey} {et~al.}(1994){Dutrey}, {Guilloteau}, \& {Simon}}]{dutrey94}
{Dutrey}, A., {Guilloteau}, S., \& {Simon}, M. 1994, \aap, 286, 149

\bibitem[{{Furlan} {et~al.}(2006){Furlan}, {Hartmann}, {Calvet}, {D'Alessio},
  {Franco-Hern{\'a}ndez}, {Forrest}, {Watson}, {Uchida}, {Sargent}, {Green},
  {Keller}, \& {Herter}}]{furlan06}
{Furlan}, E., {Hartmann}, L., {Calvet}, N., {D'Alessio}, P.,
  {Franco-Hern{\'a}ndez}, R., {Forrest}, W.~J., {Watson}, D.~M., {Uchida},
  K.~I., {Sargent}, B., {Green}, J.~D., {Keller}, L.~D., \& {Herter}, T.~L.
  2006, \apjs, 165, 568

\bibitem[{{Ghez} {et~al.}(1993){Ghez}, {Neugebauer}, \& {Matthews}}]{ghez93}
{Ghez}, A.~M., {Neugebauer}, G., \& {Matthews}, K. 1993, \aj, 106, 2005

\bibitem[{{Gordon} {et~al.}(2005){Gordon}, {Rieke}, {Engelbracht}, {Muzerolle},
  {Stansberry}, {Misselt}, {Morrison}, {Cadien}, {Young}, {Dole}, {Kelly},
  {Alonso-Herrero}, {Egami}, {Su}, {Papovich}, {Smith}, {Hines}, {Rieke},
  {Blaylock}, {P{\'e}rez-Gonz{\'a}lez}, {Le Floc'h}, {Hinz}, {Latter},
  {Hesselroth}, {Frayer}, {Noriega-Crespo}, {Masci}, {Padgett}, {Smylie}, \&
  {Haegel}}]{gordon05}
{Gordon}, K.~D., {Rieke}, G.~H., {Engelbracht}, C.~W., {Muzerolle}, J.,
  {Stansberry}, J.~A., {Misselt}, K.~A., {Morrison}, J.~E., {Cadien}, J.,
  {Young}, E.~T., {Dole}, H., {Kelly}, D.~M., {Alonso-Herrero}, A., {Egami},
  E., {Su}, K.~Y.~L., {Papovich}, C., {Smith}, P.~S., {Hines}, D.~C., {Rieke},
  M.~J., {Blaylock}, M., {P{\'e}rez-Gonz{\'a}lez}, P.~G., {Le Floc'h}, E.,
  {Hinz}, J.~L., {Latter}, W.~B., {Hesselroth}, T., {Frayer}, D.~T.,
  {Noriega-Crespo}, A., {Masci}, F.~J., {Padgett}, D.~L., {Smylie}, M.~P., \&
  {Haegel}, N.~M. 2005, \pasp, 117, 503

\bibitem[{{Gray}(1992)}]{gray92}
{Gray}, D.~F. 1992, {The observation and analysis of stellar photospheres.}
  (Camb.~Astrophys.~Ser., Vol.~20,)

\bibitem[{{Haisch} {et~al.}(2001){Haisch}, {Lada}, \& {Lada}}]{haisch01}
{Haisch}, Jr., K.~E., {Lada}, E.~A., \& {Lada}, C.~J. 2001, \apjl, 553, L153

\bibitem[{{Hartigan} \& {Kenyon}(2003)}]{hartigan03}
{Hartigan}, P., \& {Kenyon}, S.~J. 2003, \apj, 583, 334

\bibitem[{{Higdon} {et~al.}(2004){Higdon}, {Devost}, {Higdon}, {Brandl},
  {Houck}, {Hall}, {Barry}, {Charmandaris}, {Smith}, {Sloan}, \&
  {Green}}]{higdon04}
{Higdon}, S.~J.~U., {Devost}, D., {Higdon}, J.~L., {Brandl}, B.~R., {Houck},
  J.~R., {Hall}, P., {Barry}, D., {Charmandaris}, V., {Smith}, J.~D.~T.,
  {Sloan}, G.~C., \& {Green}, J. 2004, \pasp, 116, 975

\bibitem[{{Houk} \& {Cowley}(1975)}]{houk75}
{Houk}, N., \& {Cowley}, A.~P. 1975, {University of Michigan Catalogue of
  two-dimensional spectral types for the HD stars. Volume I. Declinations -90
  to -53.}, ed. N.~{Houk} \& A.~P. {Cowley}

\bibitem[{{Jensen} {et~al.}(2007){Jensen}, {Dhital}, {Stassun}, {Patience},
  {Herbst}, {Walter}, {Simon}, \& {Basri}}]{jensen07}
{Jensen}, E.~L.~N., {Dhital}, S., {Stassun}, K.~G., {Patience}, J., {Herbst},
  W., {Walter}, F.~M., {Simon}, M., \& {Basri}, G. 2007, \aj, 134, 241

\bibitem[{{Jensen} \& {Mathieu}(1997)}]{jensen97}
{Jensen}, E.~L.~N., \& {Mathieu}, R.~D. 1997, \aj, 114, 301

\bibitem[{{Jensen} {et~al.}(1996){Jensen}, {Mathieu}, \& {Fuller}}]{jensen96}
---. 1996, \apj, 458, 312

\bibitem[{{Jura} {et~al.}(1995){Jura}, {Ghez}, {White}, {McCarthy}, {Smith}, \&
  {Martin}}]{jura95}
{Jura}, M., {Ghez}, A.~M., {White}, R.~J., {McCarthy}, D.~W., {Smith}, R.~C.,
  \& {Martin}, P.~G. 1995, \apj, 445, 451

\bibitem[{{Kelson}(2003)}]{kelson03}
{Kelson}, D.~D. 2003, \pasp, 115, 688

\bibitem[{{Kelson} {et~al.}(2006){Kelson}, {Illingworth}, {Franx}, \& {van
  Dokkum}}]{kelson06}
{Kelson}, D.~D., {Illingworth}, G.~D., {Franx}, M., \& {van Dokkum}, P.~G.
  2006, \apj, 653, 159

\bibitem[{{Kelson} {et~al.}(2000){Kelson}, {Illingworth}, {van Dokkum}, \&
  {Franx}}]{kelson00}
{Kelson}, D.~D., {Illingworth}, G.~D., {van Dokkum}, P.~G., \& {Franx}, M.
  2000, \apj, 531, 159

\bibitem[{{Lada}(2006)}]{lada06}
{Lada}, C.~J. 2006, \apjl, 640, L63

\bibitem[{{Muzerolle} {et~al.}(2001){Muzerolle}, {Calvet}, \&
  {Hartmann}}]{muzerolle01}
{Muzerolle}, J., {Calvet}, N., \& {Hartmann}, L. 2001, \apj, 550, 944

\bibitem[{{Muzerolle} {et~al.}(2004){Muzerolle}, {D'Alessio}, {Calvet}, \&
  {Hartmann}}]{muzerolle04}
{Muzerolle}, J., {D'Alessio}, P., {Calvet}, N., \& {Hartmann}, L. 2004, \apj,
  617, 406

\bibitem[{{Palla} \& {Stahler}(2001)}]{palla01}
{Palla}, F., \& {Stahler}, S.~W. 2001, \apj, 553, 299

\bibitem[{{Pollack} {et~al.}(1994){Pollack}, {Hollenbach}, {Beckwith},
  {Simonelli}, {Roush}, \& {Fong}}]{pollack94}
{Pollack}, J.~B., {Hollenbach}, D., {Beckwith}, S., {Simonelli}, D.~P.,
  {Roush}, T., \& {Fong}, W. 1994, \apj, 421, 615

\bibitem[{{Preibisch} \& {Mamajek}(2008)}]{preibisch08}
{Preibisch}, T., \& {Mamajek}, E. 2008, {The Nearest OB Association:
  Scorpius-Centaurus (Sco OB2)}, ed. {Reipurth, B.}, 235--+

\bibitem[{{Quillen} {et~al.}(2004){Quillen}, {Blackman}, {Frank}, \&
  {Varni{\`e}re}}]{quillen04}
{Quillen}, A.~C., {Blackman}, E.~G., {Frank}, A., \& {Varni{\`e}re}, P. 2004,
  \apjl, 612, L137

\bibitem[{{Siess} {et~al.}(2000){Siess}, {Dufour}, \& {Forestini}}]{siess2000}
{Siess}, L., {Dufour}, E., \& {Forestini}, M. 2000, \aap, 358, 593

\bibitem[{{Simon} {et~al.}(1995){Simon}, {Ghez}, {Leinert}, {Cassar}, {Chen},
  {Howell}, {Jameson}, {Matthews}, {Neugebauer}, \& {Richichi}}]{simon95}
{Simon}, M., {Ghez}, A.~M., {Leinert}, C., {Cassar}, L., {Chen}, W.~P.,
  {Howell}, R.~R., {Jameson}, R.~F., {Matthews}, K., {Neugebauer}, G., \&
  {Richichi}, A. 1995, \apj, 443, 625

\bibitem[{{van Leeuwen}(2007)}]{vanleeuwen07}
{van Leeuwen}, F. 2007, \aap, 474, 653

\bibitem[{{White} \& {Basri}(2003)}]{white03}
{White}, R.~J., \& {Basri}, G. 2003, \apj, 582, 1109

\end{thebibliography}
\end{document}